# An Infrastructure for Software Release Analysis through Provenance Graphs

Felipe Curty[1], Troy Kohwalter[1,] Vanessa Braganholo[1], Leonardo Murta[1]

[1]Instituto de Computação – Universidade Federal Fluminense (UFF)

`felipecrp@id.uff.br, {tkohwalter,vanessa,leomurta}@ic.uff.br`

***Abstract.*** *Nowadays, quickly evolving and delivering software through a continuous delivery process is a competitive advantage and a way to keep software updated in response to the frequent changes in customers' requirements. However, the faster the software release cycle, the more challenging to track software evolution. In this paper, we propose Releasy, a tool that aims at supporting projects that use continuous delivery by generating and reporting their release provenance. The provenance generated by Releasy allows graphical visualization of the software evolution and supports queries to discover implicit information, such as the implemented features of each release and the involved developers. We also show in this paper a preliminary evaluation of Releasy in action, generating the changelog of an open source project with the provenance collected by our tool.*

## 1. Introduction

Continuous delivery aims at producing high-quality software in short release cycles and enables organizations to quickly, reliably, and efficiently evolve software, which may become an advantage over competitors (Chen, 2015). Ideally, continuous delivery enables software to be released whenever it is needed – it could be weekly, daily, or even after each commit (Neely and Stolt, 2013).

When continuous delivery is in place, tracking each release becomes a challenge due to the increasing amount of releases. Accurately knowing which issues were delivered in a specific release, the sequence of commits that were performed to implement such issues, and the developers that were in charge of such implementation is paramount for management and maintenance. The inability to change a software due to the lack of visibility over its releases may lead to losing business opportunities (Bennett and Rajlich, 2000). Hence, tracking software evolution is important, but becomes difficult in continuous delivery since the software can be released often.

Related work (Amorim Pereira and Schots, 2011; Bhattacharya et al., 2012; D'Ambros et al., 2008; De Nies et al., 2013; German and Hindle, 2006) has investigated software evolution through the use of software trails, which are the information left behind by contributors during the development process. However, only D'Ambros *et al.* (2008) handle release information. Nevertheless it is not suited to detect the issues implemented in each release and, therefore, cannot be used to generate the software changelog.

In this paper, we propose Releasy, a tool that collects provenance data from releases by parsing the software version control and issue tracking systems. Provenance describes the history of an artifact, comprising all information about the origin and derivations of that artifact (Groth and Moreau, 2013). For a given release, our tool can identify: (1) the previous release, (2) the features that were delivered in the release under analysis, (3) the commits that implemented each feature, and (4) the developers that authored the commits. Releasy can also export the collected provenance using the PROV-N notation (Groth and Moreau, 2013), which enables using the release provenance data for graphical visualizations and queries in any of the existing provenance visualization tools.

Releasy can be handy for both the practitioners and researchers. We envision the use of Releasy in a daily basis by practitioners that want to automate the generation of the public changelog of a project, to find the developers that are the most suited to fix bugs in a particular feature, and to understand the differences between releases. On the other hand, we also envision the use of Releasy by researchers that need to mine software archives and want to integrate their own mined data with release information. One could easily programmatically access this information and integrate it with other data because Releasy provides all data output in the PROV-N notation.

We used Releasy to track the releases of the *Homebrew* open source project as a method to evaluate our approach. We observed from the obtained results that our tool is capable of reconstructing the software changelog and showing accurate information about who contribute to each feature.

This paper comprises five sections besides this introduction. In Section 2, we describe our approach to collect release provenance. In Section 3, we describe Releasy, detailing how it works. In Section 4, we evaluate Releasy using the *Homebrew* open source project. In Section 5, we contrast and compare Releasy with some related work. Finally, in Section 6, we conclude the paper, discussing some future work related to our research.

## 2. Release Provenance

Our approach collects software trails from version control and issue tracking systems to build a provenance graph that helps understand the software releases. As shown in Figure 1, Releasy manages information about developers, commits, tags, and issues. The white boxes represent information gathered from the version control system, i.e., developers, commits, and tags; the gray box represents information gathered from the issue tracking system, i.e., issues; and the relationship between commits and issues is inferred by our approach. We formally define the analyzed project as $p = (D, C, I, T)$, where $D$ is the set of developers, $C$ is the set of commits, $I$ is the set of issues, and $T$ is the set of tags.

The commits represent every change made to the software and are organized in a directed acyclic graph. Figure 2 shows an example of such graph, where the commits are represented by rounded boxes (e.g., $c_{01cf}$); tags are represented by pointing boxes (e.g., $t_{3.0.15}$) and denote important software states in its history, such as releases; issues are represented by diamonds (e.g., $i_1$) and denote the change requests addressed by commits; and arrows represent the parent relationship between two commits, that is, the parent commit that is reachable from its children (e.g., $c_{a20c}$ is reachable from $c_{bc12}$ and $c_{ak4s}$).

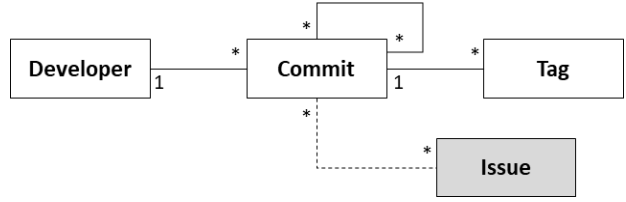

**Figure 1: Release provenance metamodel with version control system information in white and issue tracking system information in gray.**

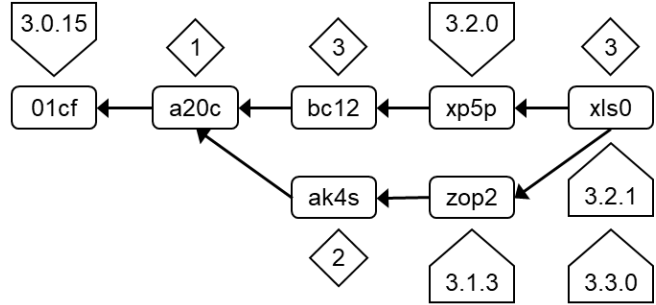

**Figure 2. A release provenance graph with commits (rounded boxes), issues (diamonds), and releases (numbered pointing boxes).**

We define each commit as $c_i = (P_i, a_i, I_i, T_i)$, where $P_i \subseteq C$ is the set of parents of the commit $c_i$, $a_i \in D$ is the author of the commit, $I_i \in I$ is the set of issues addressed by this commit, and $T_i \in T$ is the set of tags that point to the commit. From a given commit $c_i$, we form its history $H_i$ including the commit itself and all the reachable commits, recursively. This is defined as $H_i = \{c_i\} \cup \bigcup_{c_k \in P_i} H_k$. In our example, shown in Figure 2, the history of $c_{bc12}$ is $H_{bc12} = \{c_{bc12}\} \cup H_{a20c} = \{c_{bc12}, c_{a20c}, c_{01cf}\}$ and the history of $c_{xls0}$ is $H_{xls0} = \{c_{xls0}\} \cup H_{xp5p} \cup H_{zop2} = \{c_{xls0}, c_{xp5p}, c_{bc12}, c_{a20c}, c_{01cf}, c_{zop2}, c_{ak4s}\}$.

Our approach uses tags to represent releases (we provide more details in Section 3.3). Thus, we define the commit history $CH_j$ of tag $t_j$ as $CH_j = \{c \in C | c \in H_i \land t_j \in T_i\}$. We can also define all issue history $IH_j$ of tag $t_j$ as $IH_j = \bigcup_{c_i \in CH_j} I_i$, e.g., $IH_{3.2.1} = \{i_1, i_2, i_3\}$. Then, we can compare two tags in terms of released commits, e.g., $CH_{3.2.0} \setminus CH_{3.0.15} = \{c_{xp5p}, c_{bc12}, c_{a20c}\}$ or in terms of released issues, e.g., $IH_{3.2.0} \setminus IH_{3.0.15} = \{i_1, i_3\}$, where \ denotes set difference.

However, we are also interested in detecting commits and issues released by a tag $t_j$ that were not released by previous tags. Therefore, we define the tag history $TH_j = \{t_k \in T | CH_k \subset CH_j\}$, which represents the set of tags reachable from the tag $t_j$, e.g., $TH_{3.2.1} = \{t_{3.20}, t_{3.1.3}, t_{3.015}\}$. Then, we define the commits released by the tag $t_j$ that are not reachable by any previous tag as $CR_j = CH_j \setminus \bigcup_{t_k \in TH_j} CH_k$, e.g., $CR_{3.2.1} = \{c_{xls0}\}$. We can also define the issues released exclusively by the tag $t_j$ as $IR_j = \bigcup_{c_i \in CR_j} I_i$, e.g., $IR_{3.2.1} = $

$\{i_3\}$. Finally, we can identify the issues that were reworked across releases, e.g., $IR_{3.2.1} \cap IR_{3.2.0} = \{i_3\}$.

## 3. Releasy

We developed Releasy – a python tool with the primary goal to parse software trails and collect provenance from software releases. Releasy currently supports Git repositories and GitHub Issues. It offers the following features:

1. Project overview – we show the last release, the total number of releases, the number of commits, the number of developers, the number of issues related to commits, and enumerate the developers (Figure 3).
2. Release information – we show information about a specific release, including the reachable releases (the base releases), the date the release was made, the number of commits, the developers that contributed to the release, and the implemented issues (classified as features or bugfixes) (Figure 4).
3. PROV-N export – we export the release provenance graph according to the PROV-N notation, so other tools can import the data and provide additional queries and graphical visualization of the release (Figure 5).

Users can adopt Releasy by clonning the Git repository of the target project and configuring the tool with the issue tracking system URL.

### 3.1. Extracting Information from Git and GitHub

Releasy parses the Git repository using the `git log --reverse --all` command, which provides the full history of the software project, from the first to the last commit. Currently, it tracks the following information: hash (id), message, author, committer names and e-mails, and tags. With this information, Releasy populates the white boxes of the metamodel shown in Figure 1.

It also parses the GitHub Issues repository through its open REST API, which provides information about the issues related to the software. Currently, Releasy tracks the following information: id, author, subject, dates of creation and closure, and labels. With this information, it populates the gray box of the metamodel shown in Figure 1.

### 3.2. Handling Issues

While commits store all the software changes, it does not natively provide information about the implemented issues. Fortunately, many projects usually refer to the issue ID in the commit message. On projects with this behavior, it is possible to link the commit with a specific issue. For example, the message "Implements feature #1" says that this commit is related to issue with ID 1. Thus, we can associate commits and issues in our metamodel (Figure 1) by searching for the following regular expression in the commit message: "`^.*#([0-9]+).*`".

In general, issues can represent any change request, such as features and bugfixes. Though there is no direct way to identify which is the goal of an issue, we try to infer it by examining its labels on the issue tracker. Currently, we check the labels assigned to each issue and if the regular expression "`^bug.*$`" matches, we consider the issue a bugfix. Otherwise, we consider it a feature. If the project uses a different convention, then the regular expression can be adapted.

### 3.3. Handling Releases

Our approach uses tags to identify releases, which is a common pattern among software projects. However, some tags may not represent releases, i.e., waypoints to enable developers to recover a given development stage quickly. Our approach uses name convention to identify the tags related to releases. We use the regular expression "`^v?[0-9]+\.[0-9]+\.[0-9]+(-.+)?$`" to match releases. This pattern is complient with the semantic versioning[1] notation, e.g., 1.0.0 and 1.1.0-beta. Although this release name convention is vastly adopted, if the project uses a different convention, the regular expression can be adapted.

### 3.4. Data Export

Releasy can export the release provenance graph according to the PROV-N notation (Groth and Moreau, 2013). Through PROV-N, other tools can provide additional features based on the release information, such as queries and graph visualization. An example of a graph visualization is shown in Figure 5 using Prov Viewer (Kohwalter et al., 2016), which is a graph-based visualization tool for interactive exploration of provenance data.

### 4. Evaluation

We used Releasy to generate release information of *Homebrew*[2], an open source software to manage package installation in MacOS. It enables users to install and uninstall software packages on their operation system.

First, we generated the project overview, shown in Figure 3. Releasy discovered 86 releases on *Homebrew*. The last release is indicated by tag $t_{1.6.7}$. Also, the project has 15,806 commits, 694 developers, and 2,282 issues linked in the commits. In fact, the project has more issues in its issue tracking system, but Releasy only shows those that are linked to commits. Finally, Releasy lists all the project developers.

```
Project Overview
 - 1.6.7 is the last of 86 releases
 - 15806 commits made by 694 developers
 - 2282 issues linked

Developers
 - Max Howell <...@...hylblue.com>
 - Adam Vandenberg <...@...il.com>
 [...]
```

**Figure 3. Project Overview**

Then, we performed queries for specific releases. Figure 4 shows information about the release tagged as $t_{1.6.5}$. Releasy shows that the release was created on 05/25/2018 and was based on the release tagged as $t_{1.6.4}$. It also identifies 10 developers that contributed to that release. Then, Releasy listed all the 16 issues implemented in that release.

---

[1] https://semver.org/

[2] https://github.com/Homebrew/brew

```
Information about release 1.6.5
Based on: 1.6.4
Date: 2018-05-25 18:31:19+02:00
Commits: 79
Authors:
  - Markus Reiter <...@...ermark.us>
  - Gautham Goli <...@...l.com>
  [...]

Issues:
  - 4210: Update Homebrew-Cask references.
  - 4209: Reset `repo_var` so it actually is re-computed.
  - 4195: Activate Homebrew-Cask tap migration.Based on: 4.0.0-migration 3.3.7
  [...]
```

**Figure 4. Release 4.0.0 Information**

Finally, we exported a release to PROV-N and then generated the graphical visualization with Prov Viewer. We chose to present part of the release tagged as $t_{1.5.12}$ because it would better fit in the paper as an example. The result is shown in Figure 5. The release was based on the released tagged as $t_{1.5.11}$. We represented three commits: $\{c_{a264}\}$ was developed by $a_{buck}$ and $\{c_{2876}, c_{3cd5}\}$ were developed by $a_{me}$. The commits $\{c_{a264}, c_{2876}\}$ where delivered directly to the release and the commit $\{c_{3cd5}\}$ implemented issue $i_{3821}$, which was delivered to the release.

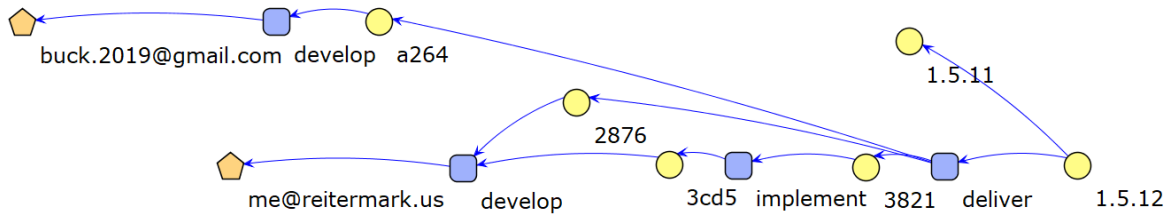

**Figure 5. Release provenance graph of release 1.5.12 of *Homebrew* visualized through Prov Viewer**

## 5. Related Work

During our research, we found other work that aims at supporting developers to track and understand software evolution through software trails. Table 1 compares them using the following information: supported version control system (VCS) and issue tracking system (ITS); and capability to handle release information (Rel.), to export provenance (Prov.), and generate software visualization (Vis.).

D'Ambros et al. (2008) is the only related work that handles release information. Their approach can analyze software evolution and collect metrics, such as the major developer and change impact. Thus, using these metrics, they can compare releases. However, they do not provide support for identifying issues released in a particular release, nor compare releases in terms of issues. Therefore, they are unable to build the software changelog. Nonetheless, in the future, their ideas can be integrated with our tool.

**Table 1. Related work of Releasy, comparing capabilities to generate PROV, build information about releases, and generate software visualization.**

| Authors | Summary | VCS | ITS | Rel. | Prov. | Vis. |
|---|---|---|---|---|---|---|
| German and Hindle (2006) | softChange is a tool to summarize, browse, and visualize the evolution of a project through its software trails. | CVS | Bugzilla | - | - | Yes |
| D'Ambros *et al.* (2008) | Present an approach populating data into a Release History Database (RHDB). | CVS | Bugzilla | Yes | - | Yes |
| Pereira and Schots (2011) | GraphVCS is a tool that enables visualization and searches through software history. | SVN | - | - | - | Yes |
| Bhattacharya *et al.* (2012) | Present an approach to characterize software evolution through graphs. | ?[3] | | - | - | - |
| De Nies *et al.* (2013) | Git2PROV is a tool to export Git history provenance in one of the following notations PROV-JSON, PROV-N, PROV-O, and SVG. | Git | - | - | Yes | Yes |
| Releasy | | Git | GitHub Issues | Yes | Yes | Yes |

## 6. Conclusion and Future Work

In this paper we presented an approach to collect software trails and generate the release provenance graph, which enables detection of the software releases, its issues, and its developers. Thus, our approach helps developers understanding the software evolution concerned to releases, which is essential in continuous delivery projects, i.e., projects that can release at any time.

We developed Releasy[4], a tool that implements our approach and enables querying information about releases. Besides that, the tool allows developers to export the collected provenance in the PROV-N notation, which enables this information to be handled by other tools compliant to PROV-N. In this paper, we use the exported provenance to generate a graphical visualization of the release in Prov Viewer.

We intend to use the exported information generated by Releasy to allow queries that exploit the inference capabilities of PROV. This would enable developers to write their own queries that traverse the commit history and infer the required information. This could significantly increase the comprehension capability of developers and end users regarding the software release history.

Finally, we have performed a preliminary evaluation on Releasy. The results suggest that Releasy can be used to track software evolution and build the software changelog. However, our tool still depends on patterns, e.g., the issue id on commit messages, which could be a problem in projects that developers do not follow such patterns.

---

[3] The authors did not specify the supported version control system and issue tracking system.
[4] Available at https://github.com/gems-uff/releasy.

In the future, we plan to evaluate Releasy in other projects by contrasting its output with other manually generated software changelogs. This way, we can check whether the quality of Releasy output is enough to replace the manual generation of release changelogs.